\documentclass[letterpaper]{article} 
\usepackage{aaai2026}  
\usepackage{times}  
\usepackage{helvet}  
\usepackage{courier}  
\usepackage[hyphens]{url}  
\usepackage{graphicx} 
\usepackage{natbib}  
\usepackage{caption} 
\usepackage{algorithm}
\usepackage{algorithmic}

\usepackage{newfloat}
\usepackage{listings}
\DeclareCaptionStyle{ruled}{labelfont=normalfont,labelsep=colon,strut=off} 
\floatstyle{ruled}
\newfloat{listing}{tb}{lst}{}
\floatname{listing}{Listing}
\title{AI Content Moderation in Therapy Conversations}

\author{
    Jiwon Kim,
    Claire Wang,
    Taeung Yoon,
    Sabelle Huang,
    Koustuv Saha
}

\affiliations{
    University of Illinois Urbana-Champaign, U.S.\\
    \{jiwonk7, claire46, tuyoon2, sabelle2, ksaha2\}@illinois.edu
}

\begin{document}

\maketitle

\begin{abstract}
  Large language models (LLMs) are increasingly being used for emotional support. They are also being developed for formal therapy purposes. However, LLMs like ChaptGPT or Llama are often developed with content moderation guardrails that prevent them from discussing sensitive subjects with users for both liability and safety purposes, and this inability to broach these subjects may affect their capacity as therapists. In this study, we perform an algorithm audit on three state-of-the-art moderation systems (OpenAI's moderation endpoint, Meta's Llama Guard, and Google's Shield Gemma) to investigate the extent to which these systems flag the content of real-life therapy sessions as undesirable. Our results raise implications for the limitations that users and organizations may encounter when designing LLMs to play the part of therapist. 
\end{abstract}

\section{Introduction}
\label{sec:intro}

Mental health remains a pressing public health challenge. In the U.S. alone, more than one in five adults live with a mental illness, and many face persistent barriers to accessing timely and affordable care~\cite{mental-illness}. At the same time, loneliness and emotional isolation have been linked to a range of serious health risks, underscoring the broader importance of accessible forms of support~\cite{eClinicalMedicine_2023}. 
As a result, people increasingly turn to online and digital systems---including online forums, chatbots, and large language model (LLM)-based agents---for emotional support, advice, and companionship~\cite{sharma2024facilitating,de2023benefits}.

To assuage their feelings of isolation, users are employing platforms such as Replika and ChatGPT to serve as emotional supports and even therapists~\cite{zhangreplika,yuan2026mental}. There has also been a growing area of research around designing LLM-based conversational agents to fulfill these roles~\cite{fang2025practicingstressreliefeveryday, 10.1145/3712299,kim2026pair}. Many of these agents, such as ChatGPT, Gemini, and Llama, are designed with content moderation guardrails~\cite{markov2023holisticapproachundesiredcontent} that prevent the conversational agent from broaching subjects that may be considered sensitive or explicit, to prevent the agent from causing harm to the user. 
Content moderation has become a fundamental aspect of video-sharing platforms, social media sites, and online forums, to facilitate productive discussions and prevent unchecked vitriol toward other users. In the case of conversational agents, many more sensitive topics that are discussed in human-to-human therapy sessions, such as suicidal ideation or thoughts of self-harm,  may be flagged by the content moderation system as inappropriate content, thereby preventing the conversation from continuing. In real-life therapy sessions, these topics are integral to discuss so that they can be parsed and processed in a healthy, controlled environment. However, this presents a possible constraint of these conversational agents in their ability to provide mental health support: if they cannot discuss these topics with users, how well can they take on roles to support their mental wellbeing? In this paper, we investigate these limitations. 

\textbf{RQ: How effectively do existing LLM moderation systems distinguish between appropriate and inappropriate content in therapist–client conversations?}


To address this question, we adopt Mahomed et. al's methodology for auditing ChatGPT's content moderation endpoint \cite{mahomed2024audit}. Building off their pipeline, we annotate a dataset of 154 high quality and 104 low quality therapy transcripts \cite{perez2019makes} to provide a ground truth for our audit using OpenAI's content moderation guidelines. Along with probing OpenAI's content moderation endpoint to assess the alignment of its flagging of violating content against the human-annotated labels, we also compare the content moderation endpoint's performance against other state-of-the-art content moderation systems, namely Meta's Llama Guard 3 \cite{llamaguard3} and Google's Shield Gemma \cite{shieldgemma}. 

We find that all three content moderation systems have low recall, with Omni Moderation at 0.09, Shield Gemma at 0.03, and Llama Guard at 0.34, and similar balanced accuracy scores of 0.53 (Omni Moderation), 0.51 (Shield Gemma), and 0.51 (Llama Guard). These results, especially the low recall across the board, suggest that these content moderation systems underflag the harms in mental health-related conversations, and while they may be able to carry out a conversation on mental health issues, they cannot identify the potential harms in those conversations, and thus may continue them past a safe context with vulnerable users. This lack of sensitivity to such harms suggests that these conversational agents are not yet ready to be deployed in mental health settings. 

Finally, we fine-tune a Mistral model using the CounselBench dataset \cite{li2025counselbench} to identify potentially harmful language in AI conversational agents. The fine-tuned Mistral model achieved a recall of 0.51 and a balanced accuracy of 0.58, an improvement from the previously assessed content moderation systems. While these scores does suggest that smaller, fine-tuned models may supply an alternative to these general-purpose content moderation systems, we find that even these fine-tuned models miss a large swathe of potentially harmful content, and are thus not yet ready to deploy in the mental health context. This provides additional insights into future attempts to make mental health resources more accessible. 

\section{Related Work}
\label{sec:related}

Mental health issues such as depression, anxiety, and suicidal thoughts continue to grow worldwide, while access to qualified mental health professionals remains severely limited~\cite{pham2022ai,wainberg2017challenges}. This imbalance, along with the persistent stigma surrounding mental illness, has encouraged individuals to seek help in digital environments. Online forums, peer-support platforms, and mental health communities have become important alternatives where people can express emotions, share experiences, and receive social validation beyond traditional therapy settings~\cite{dechoudhury2014reddit}.

With the emergence of large language models (LLMs) and generative AI, conversational agents have begun to occupy a similar role---offering continuous, personalized, and low-barrier emotional support~\cite{chen2020migrants, saha2025alzheimers,yoo2025values}. These systems have sparked interest in their potential to supplement or even replace aspects of human counseling. Yet, unlike human therapists, LLMs are heavily regulated by built-in content moderation pipelines and ``guardrails'' that restrict discussion of sensitive or high-risk topics such as self-harm, trauma, and abuse~\cite{mahomed2024audit}. While these restrictions are crucial for safety and compliance, they can also prevent models from responding to users’ genuine emotional needs, particularly in situations that require acknowledgment of distress or exploration of difficult emotions~\cite{li2025counselbench}.

Recent research has begun to examine how these moderation systems affect user experience and therapeutic quality. Studies suggest that LLMs may produce overly cautious or dismissive replies, interrupting empathic engagement or therapeutic rapport~\cite{kang2024harms}. Other works raise ethical concerns regarding the opaque decision-making of moderation algorithms and the tension between user autonomy, liability avoidance, and emotional responsiveness~\cite{shi2026mapping}. 
Recent research has also explored whether LLMs themselves can assist human moderation workflows, pointing to their promise as flexible moderation aids while also raising questions about reliability, domain sensitivity, and deployment context~\cite{kolla2024llm,kumar2024watch,SLMs}.
In parallel, recent algorithm audit literature has provided systematic frameworks for identifying algorithmic bias and distortion across domains such as search and advertising~\cite{bandy2021audits}.

Building on this line of inquiry, our study explores how ChatGPT’s content moderation system interprets and flags authentic therapy transcripts, aiming to reveal how safety guardrails shape---and sometimes constrain---the communicative capacity of LLMs designed for mental health support.

\section{Data and Methods}
\label{sec:study}
Our study aims to answer whether large language models (LLMs) triggered safety guardrails when processing therapy conversations between human therapists and patients, particularly when sensitive topics were discussed. To address this question, we analyzed the outputs of the OpenAI Moderation model (omni-moderation) \cite{omni-moderation} across therapy transcriptions, and compare them against human-generated labels. In all, there were 154 high-quality transcripts and 104 low-quality ones. We manually annotate all transcripts to serve as the ground truth and evaluate each transcript for content flags in categories such as harassment, hate, illicit content, self-harm, sexual content, and violence, as defined by OpenAI's moderation endpoint guidelines \cite{omni-moderation}. All data was anonymized, ensuring compliance with ethical research standards and protection of individual privacy. 

\begin{figure}[t]
\centering
\fbox{%
\parbox{0.95\columnwidth}{%
\small
\textbf{T:} So Jane it sounds like you're angry with your supervisor, would I be right in saying that?\\
\textbf{C:} Well I think I'm pretty angry.\\
\textbf{T:} All right, maybe to help me understand more about that anger, I'm wondering if you could choose two or three pictures here that look like how you were angry at the time ...\\
\textbf{C:} Well she got really angry at me because on the way to my room my idiot sister got in the way, so I pushed her out of the way ...
}}
\caption{Example of a flagged conversation between a Therapist (T) and a Client (C).}
\label{fig:conversation-example}
\end{figure}

\subsection{Dataset}
The data used in this project were drawn from the counseling conversations collected by Pérez-Rosas et al. \cite{perez2019makes}, who investigated the attributes of effective counseling. The dataset is comprised of 258 chat transcripts labeled as high or low quality, assessing the counselors' behaviors through the framework of the client-centered therapy principles outlined by Miller and Rollnick \cite{miller2013motivational}. Conversations in which the counselor focused on the client’s perspective and demonstrated empathy were considered high quality, while those in which the counselor primarily provided directives or potentially harmful advice, to which the client complied, were categorized as low quality. In some low-quality conversations, therapist and client labels appeared to be swapped; however, these labeling inconsistencies did not affect the overall content of the conversation, so the original labels were preserved.  

\subsection{Manual Annotation}
Our manual annotations of the transcripts served as the ground truth for determining whether a conversation should be flagged for sensitive content. Two authors manually annotated the first 25 low-quality transcripts for the categories of harassment, hate, illicit activity, self-harm, sexual content, and violence, as defined by the OpenAI moderation endpoint \cite{omni-moderation}. To evaluate the consistency of our manual harm annotations, we computed Cohen’s kappa (\(\kappa\)) between two human annotators. Cohen’s kappa measures inter-rater agreement beyond chance for categorical labels (\(\kappa = 1\) indicates perfect agreement, while \(\kappa = 0\) indicates chance-level agreement). Across the annotated data, the two annotators achieved \(\kappa = 0.64\), suggesting substantial agreement and indicating that the annotation procedure produced reasonably consistent labels between annotators. After more extensive conversation regarding the annotation criteria, the remaining transcripts (both high and low quality) were then divided between the two annotators for individual labeling.

\subsection{Audit and Finetuning}
Using the manual annotation of the transcripts, we calculated its alignment with the outputs of OpenAI's moderation endpoint, a hosted classifier which assesses the conversation across multiple harm categories and returns both binary flags and per-category severity scores (``category\_scores''), using the categories defined above; if any category exceeds the model’s internal threshold, the conversation is marked as ``flagged.''  To broaden our analysis, we also compare content moderation performance across two other moderation systems: Meta’s Llama Guard 3 \cite{llamaguard3}, and Google's Shield Gemma \cite{shieldgemma}. All prompts and inputs are standardized to ensure each system received identical text; this ensures that the differences in outputs reflected true variations in model behavior rather than input phrasing or formatting. Figure \ref{fig:conversation-example} shows a portion of a flagged conversation. Although the APIs do not provide span-level rationales indicating the exact words that triggered a flag, a conversation is flagged when at least one category crosses the model’s decision threshold.

In addition to using existing content moderation APIs, we finetune an open-source Mistral-7B model to detect therapist responses that are overtly harmful, dismissive, aggressive, or otherwise inappropriate for mental health support contexts. Our goal is to examine whether a smaller, domain-adapted language model can more effectively identify toxic counseling behaviors compared to general-purpose moderation systems, following previous research on the effectiveness of SLMs in content moderation~\cite{SLMs,goyal2025momoe}. We train this classifier with the CounselBench dataset \cite{li2025counselbench}, a large-scale, clinically grounded benchmark for evaluating open-ended mental health question answering. This finetuned model allows us to empirically compare the effectiveness of small, specialized models against widely deployed moderation APIs in mental health settings.

\section{Results}
\label{sec:results}

Therapy conversations naturally involve sensitive topics; our goal was to assess whether the LLM detected potentially harmful content appropriately and distinguished between therapeutic discussions and genuinely unsafe material. For this study, a conversation was considered flagged if any part of the transcript triggered any one of the moderation categories described above. We hypothesized that low-quality counseling sessions would yield higher harm values than high-quality ones, as ineffective counseling could fail to mitigate distress and exacerbate sensitive topics.

Importantly, the moderation systems' outputs are nondeterministic, meaning that repeated runs on the same text can yield slightly different results. This variability arises from the model architecture, infrastructure, and ongoing updates to model weights and snapshots \cite{openai-modeloptimization}. To mitigate inconsistencies in our analysis, we ran the model on each set of transcripts multiple times, similar to the approach used by \cite{mahomed2024audit}.

\subsection{Moderation Model Results}
\label{subsec:moderation_results}
We analyzed a total of 154 high-quality and 104 low-quality conversations using three moderation models: OpenAI’s omni-moderation-latest \cite{omni-moderation}, Meta’s Llama Guard 3 \cite{llamaguard3}, and Google’s ShieldGemma \cite{shieldgemma}. Each model outputs a transcript-level safety judgment using its own taxonomy. 

Across our annotated dataset, OpenAI moderation flagged 2\% of annotations as unsafe, and ShieldGemma flagged 1\%, yielding broadly similar unsafe rates. In contrast, Llama Guard 3 flagged 28\% of annotations as unsafe. This difference is largely explained by Llama Guard 3’s inclusion of a distinct ``Specialized Advice'' category, which treats specialized financial, medical, or legal advice as unsafe content \cite{llamaguard3}. In our results, approximately 90\% of Llama Guard 3 unsafe annotations fell into this Specialized Advice category. Because the dialogues in the dataset often include therapists offering suggestions and guidance, Llama Guard 3’s taxonomy is more likely to mark clinically routine exchanges as unsafe, inflating the overall flag rate relative to the other two models.

To assess intra-model consistency, we ran each model on each conversation five times and aggregated the outputs using majority vote for the final unsafe/flagged decision. Only a small number of cases varied across runs, typically by adding or removing a single secondary category while preserving the overall safe/unsafe decision.

\label{subsec:manual_annotation}



\subsection{Finetuned Model Results}

\begin{table}[t]
\centering
\caption{Recall and balanced accuracy for detecting toxic therapist responses. The finetuned Mistral model achieves the highest performance on both metrics.}
\label{tab:toxicity_detection}
\begin{tabular}{lcc}
\textbf{Model} & \textbf{Recall} & \textbf{Balanced Acc.} \\
\hline
Omni Moderation & 0.0860 & 0.5260 \\
Shield Gemma & 0.0286 & 0.5143 \\
Llama Guard & 0.3429 & 0.5089 \\
Our Finetuned Mistral & \textbf{0.5143} & \textbf{0.5821} \\
\hline
\end{tabular}
\end{table}

Table~\ref{tab:toxicity_detection} reports recall and balanced accuracy for detecting toxic therapist responses, comparing multiple moderation systems against human-labeled ground truth. In this evaluation, responses flagged by each moderation model were aligned with annotations indicating whether a therapist's utterance was toxic or inappropriate in a counseling context. Among the evaluated systems, Omni Moderation achieves a recall of 0.0860 with a balanced accuracy of 0.5260, while Shield Gemma performs worse in recall (0.0286) with a similar balanced accuracy of 0.5143. Llama Guard improves recall to 0.3429 but maintains a comparable balanced accuracy of 0.5089. In contrast, our finetuned Mistral judge achieves the highest recall (0.5143) and balanced accuracy (0.5821), outperforming both general-purpose moderation APIs and safety-focused models such as Shield Gemma and Llama Guard.

Despite its smaller size, the finetuned Mistral model achieves the highest recall among all evaluated systems (0.5143), substantially outperforming Llama Guard (0.3429), Omni Moderation (0.0860), and Shield Gemma (0.0286). This improvement in recall indicates that our model is more effective at identifying toxic or clinically inappropriate responses, which is particularly critical in safety-sensitive mental health contexts. At the same time, the absolute recall values remain moderate across all models, suggesting that current moderation systems—whether API-based or safety-focused—fail to detect a significant portion of harmful responses. These results indicate that manually finetuning a smaller model with domain-specific data can be more effective than relying on larger, general-purpose moderation systems, and further highlight the need for additional supervision mechanisms beyond surface-level toxicity detection.

\section{Discussion and Conclusion}

There are several approaches that could be additionally explored to overcome or adjust LLM safety guardrails, such as prompt engineering, few-shot learning, or fine-tuning open-source models. Prompt engineering can be used to reframe user inputs or system instructions in ways that reduce refusal behaviors while preserving task relevance. Few-shot learning provides carefully curated examples that guide the model toward desired response styles, potentially steering it away from overly conservative safety responses. Fine-tuning open-source models enables direct modification of model behavior using domain-specific datasets, allowing practitioners to tailor safety boundaries and response characteristics more precisely. Recent research \cite{song2025refusal} has shown that machine unlearning can be used to break safety alignment by making an LLM forget its mechanisms for rejecting harmful instructions. These methods are potential directions for future work.

At the same time, directly weakening safety guardrails introduces significant risks, particularly in sensitive domains such as mental health. Rather than removing safeguards, an alternative direction is to complement existing guardrails with structured, domain-specific supervision mechanisms that enable more nuanced control over model behavior. For example, integrating rubric-based evaluation frameworks or iterative critique-and-revision pipelines may allow systems to remain aligned with high-level safety policies while improving their ability to handle complex, context-dependent interactions~\cite{kim2026pair,goel2026rubrix}. 
Exploring how such approaches can balance flexibility, safety, and clinical appropriateness remains an important direction for future research.

\label{sec:discussion}
Overall, we found that LLMs alone are not enough to fulfill the role of a therapist. In general, the models did a poor job in determining whether a conversation should be flagged for sensitive content. It is unrealistic to expect the AI to catch every coded or indirect expression, so an LLM model alone could not replace a therapist. 

However, in real-world therapy settings, people are usually more straightforward when discussing their emotions, so the risk of such bypassing language affecting AI responses is likely minimal. Still, this prompt sensitivity should be acknowledged as one of the study’s limitations.

We explored whether moderation behavior could be made more comparable across systems by testing models with other models’ taxonomies and guidelines (e.g., evaluating the same transcript with one model while providing the category definitions and decision rules from another). In practice, this did not produce meaningful or reliable results. Classifier-style moderation endpoints (e.g., omni-moderation~\cite{omni-moderation}) did not adopt external taxonomies beyond their fixed label spaces. As a result, these cross-instruction trials did not improve interpretability or alignment between models’ unsafe judgments, and we treat them as an empirical takeaway about the limits of prompt-based harmonization rather than a basis for our main analyses.

\bibliography{main}

\begin{thebibliography}{33}
\providecommand{\natexlab}[1]{#1}

\bibitem[{AI(2024)}]{llamaguard3}
AI, M. 2024.
\newblock Llama Guard 3: Safety Classifier and Moderation Model.
\newblock \url{https://www.llama.com/docs/model-cards-and-prompt-formats/llama-guard-3/}.
\newblock Accessed: 2025-10-12.

\bibitem[{Bandy(2021)}]{bandy2021audits}
Bandy, J. 2021.
\newblock Problematic Machine Behavior: A Systematic Literature Review of Algorithm Audits.
\newblock \emph{Proceedings of the ACM on Human-Computer Interaction}, 5(CSCW1): 1--34.

\bibitem[{Chen et~al.(2020)Chen, Lu, Nieminen, and Lucero}]{chen2020migrants}
Chen, Z.; Lu, Y.; Nieminen, M.~P.; and Lucero, A. 2020.
\newblock Creating a Chatbot for and with Migrants: Chatbot Personality Drives Co-Design Activities.
\newblock In \emph{Proceedings of the 2020 ACM Designing Interactive Systems Conference}, 219--230.

\bibitem[{De~Choudhury and De(2014)}]{dechoudhury2014reddit}
De~Choudhury, M.; and De, S. 2014.
\newblock Mental Health Discourse on Reddit: Self-Disclosure, Social Support, and Anonymity.
\newblock In \emph{Eighth International AAAI Conference on Weblogs and Social Media}.

\bibitem[{De~Choudhury, Pendse, and Kumar(2023)}]{de2023benefits}
De~Choudhury, M.; Pendse, S.~R.; and Kumar, N. 2023.
\newblock Benefits and harms of large language models in digital mental health.
\newblock \emph{arXiv preprint arXiv:2311.14693}.

\bibitem[{eClinicalMedicine(2023)}]{eClinicalMedicine_2023}
eClinicalMedicine. 2023.
\newblock The epidemic of loneliness.
\newblock \emph{eClinicalMedicine}, 66: 102395.

\bibitem[{Fang et~al.(2025)Fang, Chhabria, Maram, and Zhu}]{fang2025practicingstressreliefeveryday}
Fang, A.; Chhabria, H.; Maram, A.; and Zhu, H. 2025.
\newblock Practicing Stress Relief for the Everyday: Designing Social Simulation Using VR, AR, and LLMs.
\newblock arXiv:2410.01672.

\bibitem[{Goel et~al.(2026)Goel, Lee, Zhong, Rodriguez, Brown, Karkar, Yoo, and Saha}]{goel2026rubrix}
Goel, D.; Lee, J.; Zhong, Q.~J.; Rodriguez, V.~J.; Brown, D.~S.; Karkar, R.; Yoo, D.~W.; and Saha, K. 2026.
\newblock RubRIX: Rubric-Driven Risk Mitigation in Caregiver-AI Interactions.
\newblock In \emph{Findings of the Association for Computational Linguistics (ACL)}.

\bibitem[{Google(2024)}]{shieldgemma}
Google. 2024.
\newblock ShieldGemma Model Card.
\newblock \url{https://ai.google.dev/gemma/docs/shieldgemma/model_card_2}.
\newblock Accessed: 2025-10-12.

\bibitem[{Goyal et~al.(2025)Goyal, Zhan, Chen, Saha, and Chandrasekharan}]{goyal2025momoe}
Goyal, A.; Zhan, X.; Chen, Y.; Saha, K.; and Chandrasekharan, E. 2025.
\newblock Momoe: Mixture of moderation experts framework for ai-assisted online governance.
\newblock In \emph{Proceedings of the 2025 Conference on Empirical Methods in Natural Language Processing}, 12656--12671.

\bibitem[{Kang and Reynolds(2024)}]{kang2024harms}
Kang, R.~M.; and Reynolds, T.~L. 2024.
\newblock "This App Said I Had Severe Depression, and Now I Don’t Know What to Do": The Unintentional Harms of Mental Health Applications.
\newblock In \emph{Proceedings of the CHI Conference on Human Factors in Computing Systems}, 1--17.

\bibitem[{Kim et~al.(2026)Kim, Rodriguez, Yoo, Chandrasekharan, and Saha}]{kim2026pair}
Kim, J.; Rodriguez, V.~J.; Yoo, D.~W.; Chandrasekharan, E.; and Saha, K. 2026.
\newblock PAIR-SAFE: A Paired-Agent Approach for Runtime Auditing and Refining AI-Mediated Mental Health Support.
\newblock \emph{arXiv preprint arXiv:2601.12754}.

\bibitem[{Kolla et~al.(2024)Kolla, Salunkhe, Chandrasekharan, and Saha}]{kolla2024llm}
Kolla, M.; Salunkhe, S.; Chandrasekharan, E.; and Saha, K. 2024.
\newblock Llm-mod: Can large language models assist content moderation?
\newblock In \emph{Extended Abstracts of the CHI Conference on Human Factors in Computing Systems}, 1--8.

\bibitem[{Kumar, AbuHashem, and Durumeric(2024)}]{kumar2024watch}
Kumar, D.; AbuHashem, Y.~A.; and Durumeric, Z. 2024.
\newblock Watch your language: Investigating content moderation with large language models.
\newblock In \emph{Proceedings of the International AAAI Conference on Web and Social Media}, volume~18, 865--878.

\bibitem[{Li et~al.(2025{\natexlab{a}})Li, Zhu, Zhang, and Lee}]{zhangreplika}
Li, J.; Zhu, Z.; Zhang, R.; and Lee, Y.-C. 2025{\natexlab{a}}.
\newblock Exploring the Effects of Chatbot Anthropomorphism and Human Empathy on Human Prosocial Behavior Toward Chatbots.
\newblock \emph{Proc. ACM Hum.-Comput. Interact.}, 9(7).

\bibitem[{Li et~al.(2025{\natexlab{b}})Li, Yao, Bunyi, Frank, Hwang, and Liu}]{li2025counselbench}
Li, Y.; Yao, J.; Bunyi, J. B.~S.; Frank, A.~C.; Hwang, A.; and Liu, R. 2025{\natexlab{b}}.
\newblock CounselBench: A Large-Scale Expert Evaluation and Adversarial Benchmarking of Large Language Models in Mental Health Question Answering.
\newblock \emph{arXiv preprint arXiv:2506.08584}.

\bibitem[{Mahomed et~al.(2024)}]{mahomed2024audit}
Mahomed, Y.; et~al. 2024.
\newblock Auditing GPT’s Content Moderation Guardrails: Can ChatGPT Write Your Favorite TV Show?
\newblock In \emph{Proceedings of the 2024 ACM Conference on Fairness, Accountability, and Transparency (FAccT)}, 660--686. Rio de Janeiro, Brazil.

\bibitem[{Markov et~al.(2023)Markov, Zhang, Agarwal, Eloundou, Lee, Adler, Jiang, and Weng}]{markov2023holisticapproachundesiredcontent}
Markov, T.; Zhang, C.; Agarwal, S.; Eloundou, T.; Lee, T.; Adler, S.; Jiang, A.; and Weng, L. 2023.
\newblock A Holistic Approach to Undesired Content Detection in the Real World.
\newblock arXiv:2208.03274.

\bibitem[{Miller and Rollnick(2013)}]{miller2013motivational}
Miller, W.~R.; and Rollnick, S. 2013.
\newblock \emph{Motivational Interviewing: Helping People Change}.
\newblock New York, NY, USA: Guilford Press, 3rd edition.

\bibitem[{Nie et~al.(2025)Nie, Shao, Fan, Shao, You, Preindl, and Jiang}]{10.1145/3712299}
Nie, J.; Shao, H.~V.; Fan, Y.; Shao, Q.; You, H.; Preindl, M.; and Jiang, X. 2025.
\newblock LLM-based Conversational AI Therapist for Daily Functioning Screening and Psychotherapeutic Intervention via Everyday Smart Devices.
\newblock \emph{ACM Transactions on Computing for Healthcare}.
\newblock Just Accepted.

\bibitem[{{NIH}(2024)}]{mental-illness}
{NIH}. 2024.
\newblock Mental Illness.
\newblock \url{https://www.nimh.nih.gov/health/statistics/mental-illness}.
\newblock Accessed: 2025-10-11.

\bibitem[{{OpenAI}(2024)}]{openai-modeloptimization}
{OpenAI}. 2024.
\newblock Model Optimization Guide.
\newblock \url{https://platform.openai.com/docs/guides/model-optimization}.
\newblock Accessed: 2025-10-11.

\bibitem[{OpenAI(2025)}]{omni-moderation}
OpenAI. 2025.
\newblock OpenAI omni-moderation-latest Model Documentation.
\newblock \url{https://platform.openai.com/docs/models/omni-moderation-latest}.

\bibitem[{P{\'e}rez-Rosas et~al.(2019)P{\'e}rez-Rosas, Wu, Resnicow, and Mihalcea}]{perez2019makes}
P{\'e}rez-Rosas, V.; Wu, X.; Resnicow, K.; and Mihalcea, R. 2019.
\newblock What makes a good counselor? learning to distinguish between high-quality and low-quality counseling conversations.
\newblock In \emph{Proceedings of the 57th Annual Meeting of the Association for Computational Linguistics}, 926--935.

\bibitem[{Pham, Nabizadeh, and Selek(2022)}]{pham2022ai}
Pham, K.~T.; Nabizadeh, A.; and Selek, S. 2022.
\newblock Artificial Intelligence and Chatbots in Psychiatry.
\newblock \emph{Psychiatric Quarterly}, 93(1): 249--253.

\bibitem[{Saha et~al.(2025)Saha, Jain, Liu, Kaliappan, and Karkar}]{saha2025alzheimers}
Saha, K.; Jain, Y.; Liu, C.; Kaliappan, S.; and Karkar, R. 2025.
\newblock AI vs. Humans for Online Support: Comparing the Language of Responses from LLMs and Online Communities of Alzheimer's Disease.
\newblock \emph{ACM Transactions on Computing for Healthcare}.

\bibitem[{Sharma et~al.(2024)Sharma, Rushton, Lin, Nguyen, and Althoff}]{sharma2024facilitating}
Sharma, A.; Rushton, K.; Lin, I.~W.; Nguyen, T.; and Althoff, T. 2024.
\newblock Facilitating self-guided mental health interventions through human-language model interaction: A case study of cognitive restructuring.
\newblock In \emph{Proc. CHI}.

\bibitem[{Shi et~al.(2026)Shi, Yoo, Wang, Rodriguez, Karkar, and Saha}]{shi2026mapping}
Shi, J.~M.; Yoo, D.~W.; Wang, K.; Rodriguez, V.~J.; Karkar, R.; and Saha, K. 2026.
\newblock Mapping Caregiver Needs to AI Chatbot Design: Strengths and Gaps in Mental Health Support for Alzheimer's and Dementia Caregivers.
\newblock \emph{ACM Transactions on Computing for Healthcare}.

\bibitem[{Song et~al.(2025)Song, Kim, Kim, Shin, and Son}]{song2025refusal}
Song, M.; Kim, H.; Kim, J.; Shin, S.; and Son, S. 2025.
\newblock Refusal Is Not an Option: Unlearning Safety Alignment of Large Language Models.
\newblock In \emph{34th USENIX Security Symposium (USENIX Security 25)}, 319--338.

\bibitem[{Wainberg et~al.(2017)Wainberg, Scorza, Shultz, Helpman, Mootz, Johnson, Neria, Bradford, Oquendo, and Arbuckle}]{wainberg2017challenges}
Wainberg, M.~L.; Scorza, P.; Shultz, J.~M.; Helpman, L.; Mootz, J.~J.; Johnson, K.~A.; Neria, Y.; Bradford, J.-M.~E.; Oquendo, M.~A.; and Arbuckle, M.~R. 2017.
\newblock Challenges and Opportunities in Global Mental Health: A Research-to-Practice Perspective.
\newblock \emph{Current Psychiatry Reports}, 19: 1--10.

\bibitem[{Yoo et~al.(2026)Yoo, Shi, Rodriguez, and Saha}]{yoo2025values}
Yoo, D.~W.; Shi, J.~M.; Rodriguez, V.~J.; and Saha, K. 2026.
\newblock AI Chatbots for Mental Health Self-Management: Lived Experience--Centered Qualitative Study.
\newblock \emph{JMIR Mental Health}, 13: e78288.

\bibitem[{Yuan et~al.(2026)Yuan, Zhang, Aledavood, Zhang, and Saha}]{yuan2026mental}
Yuan, Y.; Zhang, J.; Aledavood, T.; Zhang, R.; and Saha, K. 2026.
\newblock Mental Health Impacts of AI Companions: Triangulating Social Media Quasi-Experiments, User Perspectives, and Relational Theory.
\newblock In \emph{Proceedings of the 2026 CHI Conference on Human Factors in Computing Systems}.

\bibitem[{Zhan et~al.(2025)Zhan, Goyal, Chen, Chandrasekharan, and Saha}]{SLMs}
Zhan, X.; Goyal, A.; Chen, Y.; Chandrasekharan, E.; and Saha, K. 2025.
\newblock SLM-mod: Small language models surpass LLMs at content moderation.
\newblock In \emph{Proc. NAACL}.

\end{thebibliography}

\end{document}